\documentclass[aps,prb,twocolumn,showpacs,amsfonts,superscriptaddress]{revtex4}

\usepackage{epsfig}
\usepackage{amsmath}
\usepackage{amssymb}
\usepackage{bm}

\newcommand{\beq}{\begin{equation}}
\newcommand{\eeq}{\end{equation}}
\newcommand{\beqarray}{\begin{eqnarray}}
\newcommand{\eeqarray}{\end{eqnarray}}

\newcommand{\eq}[1]{Eq.~(\ref{#1})} 
\newcommand{\fig}[1]{Fig.~(\ref{#1})} 
\newcommand{\pypy}{$p_y$-$p_y$~}
\newcommand{\pzpz}{$p_z$-$p_z$~}
\newcommand{\Hc}{\ensuremath{\mbox{H.c.}}} 

\allowdisplaybreaks

\begin{document}

\title{Functional superconductor interfaces from broken time-reversal
    symmetry} 
\author{P. M. R. Brydon}
\email{brydon@theory.phy.tu-dresden.de}
\affiliation{Max-Planck-Institut f\"{u}r Festk\"{o}rperforschung,
  Heisenbergstr. 1, 70569 Stuttgart, Germany}
\affiliation{Institut f\"{u}r Theoretische Physik, Technische Universit\"{a}t
  Dresden, 01062 Dresden, Germany}
\author{Christian Iniotakis}
\affiliation{Institut f\"{u}r Theoretische Physik, ETH Z\"{u}rich, CH-8093
Z\"{u}rich, Switzerland}
\author{Dirk Manske}
\affiliation{Max-Planck-Institut f\"{u}r Festk\"{o}rperforschung,
  Heisenbergstr. 1, 70569 Stuttgart, Germany}
\author{M. Sigrist} 
\affiliation{Institut f\"{u}r Theoretische Physik, ETH Z\"{u}rich, CH-8093
Z\"{u}rich, Switzerland}

\date{\today}

\begin{abstract}
The breaking of time-reversal symmetry in a triplet superconductor Josephson
junction is shown to cause a magnetic instability of the tunneling
barrier. Using a Ginzburg-Landau analysis of the free energy, we predict that
this novel functional behaviour reflects the formation of an
exotic Josephson state, distinguished by the
existence of fractional flux quanta at the barrier. The crucial role of the
orbital pairing state is demonstrated by studying complementary microscopic
models of the junction. Signatures of the magnetic instability are found in
the critical current of the junction.
\end{abstract}


\maketitle

The Josephson effect between superconductors separated by a tunneling barrier
continues to be of fundamental interest, in
particular as a phase-sensitive test of the pairing symmetry of unconventional 
superconductors,~\cite{Kashiwaya2000,Tsuei2000} and in the study of the
interplay of superconductivity with
magnetism.~\cite{Golubov2004,SCFM} Although the 
qualitative features of the Josephson effect are determined by the quantum
nature of the superconductors, the modification of the
superconducting state at the junction interface must often be included in the
proper description of the supercurrent transmission, the so-called proximity
effect.~\cite{Kashiwaya2000,Golubov2004,SCFM}
In contrast, the properties of the barrier are usually regarded as
fixed.~\cite{Kashiwaya2000} Recently,
however, it has been shown that a thin ferromagnetic layer on a  
singlet superconductor can display novel behaviour.~\cite{SCFM} In particular,
the presence of the superconductor can suppress the uniform
magnetization~\cite{Begeret2004} or 
stabilize a domain structure in the ferromagnet.~\cite{crypto} These effects
result from the competition between singlet superconductivity and
ferromagnetism, and indicate that the tunneling barrier in a
Josephson junction is not necessarily independent of the superconductors.

In this letter, we consider the possibility of using the presence of the
two superconductors to \emph{induce} a magnetic instability of a non-magnetic
tunneling barrier in a Josephson junction. In particular, by both general
phenomenological arguments and solution of specific microscopic models, we show
that such a novel functionality of the barrier can develop 
for time-reversal symmetry (TRS) breaking configurations of two spin-triplet
superconductors on either side. The Josephson coupling across the tunneling
barrier is essential to this effect, which manifests itself as an exotic
state distinguished, for example, by the existence of fractional flux quanta
at the barrier.~\cite{Sigrist1995}
Moreover, such a junction displays an anomalous
temperature dependence of the critical current. 

The intrinsic spin structure of the Cooper pairs in a triplet superconductor
(TSC) requires a quasiparticle gap with three components 
$\widetilde{\Delta}_{S_z}$ for each $z$-component of spin $S_z=-1, 0,
+1$. Each of these gaps has odd orbital parity,
i.e. $\widetilde{\Delta}_{S_z}(-{\bf{k}})=-\widetilde{\Delta}_{S_z}({\bf{k}})$. The
pairing state is conveniently 
described by the so-called ${\bf{d}}$-vector, defined in spin-space
${\bf{d}}=\tfrac{1}{2}(\widetilde{\Delta}_{-1}-\widetilde{\Delta}_{1}){\bf{x}} - 
\tfrac{i}{2}(\widetilde{\Delta}_{-1}+\widetilde{\Delta}_{1}){\bf{y}} 
+ \widetilde{\Delta}_{0}{\bf{z}}$, which also serves as the order parameter
for the TSC.
Although a multitude of different triplet pairing states are allowed by
symmetry, only a few examples of TSCs have been discovered so 
far, e.g. Sr$_2$RuO$_4$,~\cite{Maeno1994} UGe$_2$.~\cite{UGe2} In
Sr$_2$RuO$_4$ the spin pairing state has been 
identified as unitary and equal-spin-pairing,~\cite{Mackenzie2003} i.e. the
spins of the triplet Cooper pairs lie in the same plane in spin space
perpendicular to ${\bf{d}}$,
but the condensate does not have a net spin. Restricting
ourselves to such pairing states, we
write ${\bf{d}}=\widetilde{\bf{d}}e^{i\phi}$ where $\widetilde{\bf{d}}$ is a
real vector.
The vector character of the TSC order parameter provides a
novel degree of freedom in Josephson junction physics: in addition to the
phase difference between the condensates to the left and right of the barrier,
which controls the 
Josephson supercurrent as in the familiar spin-singlet case, there is also 
the mutual alignment of the left ($\widetilde{\bf{d}}_{L}$) and right
($\widetilde{\bf{d}}_R$) vectors, which controls the magnetic aspects of the
transport.~\cite{Asano2006} In particular, when
$\widetilde{\bf{d}}_{L}\times\widetilde{\bf{d}}_{R}\neq0$, a Cooper pair
tunneling across the barrier 
undergoes a reconstruction of its spin state, producing
an effective spin through the TRS breaking combination 
$\langle{\bf{S}}\rangle=i{\bf{d}}^{}_{L}\times{\bf{d}}^{\ast}_{R} +
\Hc$ at the junction interface.

The violation of TRS at the tunneling barrier
by the 
misaligned $\widetilde{\bf{d}}$-vectors allows the TSCs to
directly 
couple to a magnetization ${\bf{M}}$ of the interface. 
As we will
see, the TSCs may in fact change the electronic properties
of the interface so as to stabilize a spontaneous ferromagnetic
order. 
This novel behavior can be understood on a phenomenological level 
by a Ginzburg-Landau analysis of the
free energy. Introducing the TSC order
parameters for each side of the
interface as ${\bf{d}}_{L}=\widetilde{\bf{d}}_{L}e^{i\phi_L}$ and
${\bf{d}}_{R}=\widetilde{\bf{d}}_Re^{i\phi_R}$, we write the free energy of
the junction $F$ to lowest order in 
${\bf{M}}$ and $\widetilde{\bf{d}}_{L,R}$ as
\beq
F =
\frac{|{\bf{M}}|^2}{2\chi} -
t\widetilde{\bf{d}}_{L}\cdot\widetilde{\bf{d}}_R\cos(\phi) + 
2\gamma{\bf{M}}\cdot(\widetilde{\bf{d}}_{L}\times\widetilde{\bf{d}}_R)\sin(\phi) \label{eq:GL} 
\eeq
Here $\phi=\phi_R-\phi_L$ is the phase difference, $\chi$ denotes the
intrinsic uniform spin 
susceptibility of the barrier, and $t$ and $\gamma$ are phenomenological
parameters.  
The ground state of the junction is obtained by minimizing $F$ with respect to
${\bf{M}}$ and $\phi$. We find that the non-magnetic state of the barrier is
unstable for   
$\chi \gamma^2 | \widetilde{\bf{d}}_{L}\times\widetilde{\bf{d}}_R |^2 > t\widetilde{\bf{d}}_{L}\cdot\widetilde{\bf{d}}_R$: the coupling to
the intrinsic spin
$\langle{\bf{S}}\rangle=2\widetilde{\bf{d}}_{L}\times\widetilde{\bf{d}}_R\sin(\phi)$
of the junction instead realizes a TRS breaking state,
characterized by a non-vanishing magnetization
${\bf{M}}\parallel\widetilde{\bf{d}}_{L}\times\widetilde{\bf{d}}_R$, and a
phase 
$\phi=\phi_{\text{min}}\neq0,\pi$. The junction is then in an
exotic \emph{fractional} state,~\cite{Sigrist1995} where the magnetic barrier
is capable of carrying flux lines with non-integer multiples of the flux
quantum $\Phi_0=hc/2e$. The observation of this characteristic feature is discussed below.

\begin{figure}
  \includegraphics[width=0.8\columnwidth]{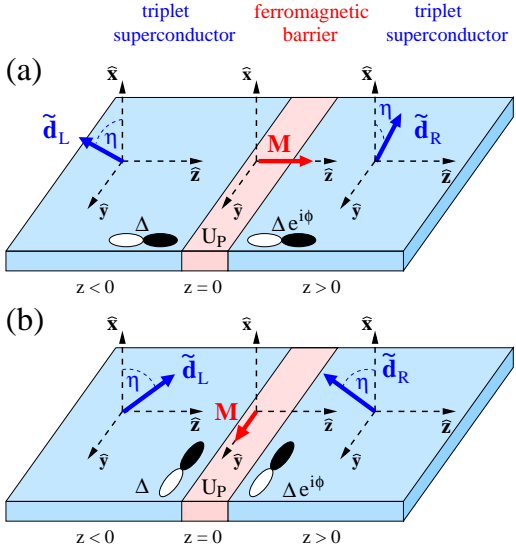}
  \caption{\label{junction} (color online) 
    (a) The \pzpz junction, with $p_{z}$-wave orbital pairing in the two 
    superconductors, $\widetilde{\bf{d}}$-vectors in the $x$-$y$ plane, and an
    induced 
    magnetic moment along the $z$-axis. (b) The \pypy junction, with
    $p_{y}$-wave orbital pairing in the two 
    superconductors, $\widetilde{\bf{d}}$-vectors in the $x$-$z$ plane, and an
    induced 
    magnetic moment along the $y$-axis.} 
\end{figure}

The magnetic instability depends essentially upon, and can be tuned by, the
mutual misalignment of the $\widetilde{\bf{d}}$-vectors of the two
TSCs. It 
is also controlled by the specific details of the junction, through the
susceptibility $\chi$, and the parameters $t$ and $\gamma$. The latter are
fixed by the orbital part of the bulk pairing state.
To elucidate the crucial role this plays
in the magnetic instability, we examine two complementary microscopic models
of the junction.
The first, shown 
in~\fig{junction}(a), has the TSCs in a $p_z$-wave orbital
state (the \pzpz junction), and the left and right 
$\widetilde{\bf{d}}$-vectors are aligned parallel to the barrier
but at an angle $2\eta$ with respect to each other. In the second model the
orbital state is $p_y$-wave (the \pypy junction), and the
$\widetilde{\bf{d}}$-vectors lie in the $x$-$z$ plane but are
again misoriented by the angle $2\eta$ [\fig{junction}(b)]. We assume 
translational invariance in the $x$-$y$ 
plane and that the bulk TSCs extend indefinitely along the
$z$-axis. Furthermore, we take the TSCs to have spatially-constant gaps, and 
the maximum gap magnitude $\Delta$ displays
weak-coupling temperature-dependence, with $T=0$ value $\Delta_{0}$. The
tunneling barrier is modeled to be of $\delta$-function width,
with normal-state height $U_{P}=Z\hbar{v_{F}}$ where $Z$ is a dimensionless
quantity and $v_{F}=\hbar{k_F}/m$ is the Fermi velocity in the bulk TSCs,
which are assumed to have spherical Fermi sufaces of radius $k_F$. If the
barrier supports a magnetic moment  
${\bf{M}}\parallel{\widetilde{\bf{d}}_{L}\times\widetilde{\bf{d}}_R}$, the 
effective barrier height for spin-$\sigma$ quasiparticles with spin parallel
to ${\bf{M}}$ in dimensionless units is $Z-\sigma{M}$ where
$M=g\mu_{B}|{\bf{M}}|/\hbar{v_F}$.  

\begin{figure}
  \includegraphics[width=\columnwidth]{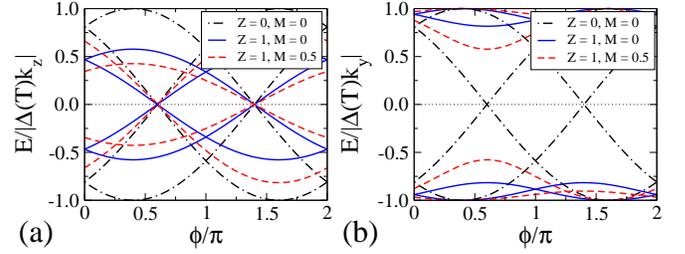}
  \caption{\label{ABS} (color online) The Andreev bound state spectrum in the
    (a) \pzpz and (b) the \pypy junction
    for $k_{z}=k_{y}=k_{F}/\sqrt{2}$. }  
\end{figure}

The electronic properties of the two junctions can be expressed entirely
in terms of the Andreev bound state (ABS) spectrum.~\cite{Kashiwaya2000} These
subgap states are localized at the barrier interface and are 
formed by multiple Andreev reflection of tunneling quasiparticles
within the barrier. We solve the
Bogoliubov-de Gennes equations~\cite{BrydonTFT2008} to obtain explicit
expressions for the ABS energies $E_{{\bf{k}},\sigma}$ in the
\pzpz junction 
\beq
E_{{\bf{k}},\sigma} =
|\Delta(T)k_{z}|\sqrt{{\cal{T}_{\sigma}({\bf{k}})}}\cos(\phi/2
-\sigma\eta) \label{eq:ABSpzpz} 
\eeq
and the \pypy junction
\beq
E_{{\bf{k}},\sigma} =
|\Delta(T)k_{y}|\sqrt{1-{\cal{T}_{\sigma}({\bf{k}})}\sin^2(\phi/2+\sigma\eta)} \label{eq:ABSpypy}
\eeq
The bound states are indexed by the component of spin
$\sigma=\pm1$ parallel to
$\widetilde{\bf{d}}_L\times\widetilde{\bf{d}}_R$, and each state has two
branches at $\pm{E}_\sigma$. The transparency of the barrier to spin-$\sigma$
quasiparticles is given by ${\cal{T}}_{\sigma}({\bf{k}})=k_{z}^2/[k_{z}^2 +
  (Z-\sigma{M})^2k_F^2]$. We plot the ABS spectrum as a
function of $\phi$ in~\fig{ABS}. Note that the bound states always intersect
the line $E=0$ in the \pzpz junction. These so-called zero energy state are
guaranteed by the arrangement of the $p$-wave orbitals, such that all
specularly-reflected quasiparticles experience a sign-change of the 
gap. Since the gap does not change sign for reflected quasiparticles in the
\pypy junction, in contrast, zero energy states  
are only found here for a perfectly transparent tunneling barrier, as is also
the case for $s$-wave superconductor junctions.~\cite{Kashiwaya2000}

\begin{figure}
  \includegraphics[width=\columnwidth]{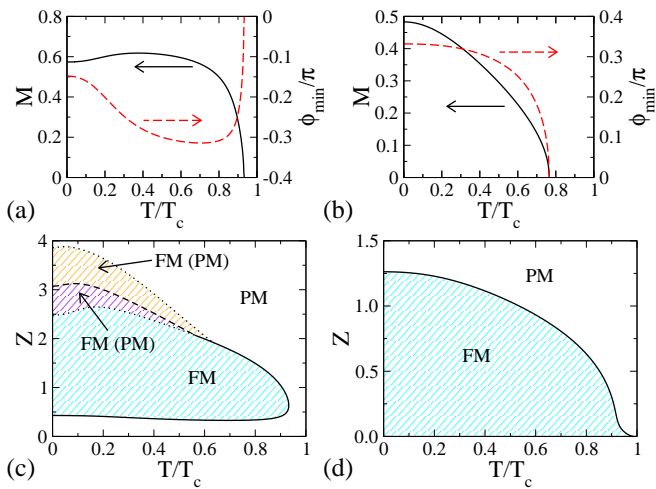}
  \caption{\label{fig3} (color online) Induced magnetic moment $M$  and stable
    phase difference $\phi_\text{min}$ as a function of reduced temperature
    for (a) the \pzpz and (b) the \pypy junction. In both
    panels we set $\eta=0.2\pi$, $Z=0.7$ and
    $\chi=20$. Magnetic phase diagram for the (c) \pzpz and (d)
    \pypy junctions as a function of $Z$ and $T$. A magnetic moment is
    stable in the region labeled FM, while the region of
    non-magnetic behaviour is denoted as PM; metastable
    states are shown in brackets. Second-order transitions are indicated by
    a solid line, first-order transitions by a dashed line, and the limits of
    the metastable states by a dotted line. $\eta$ and $\chi$ are as
    in (a) and (b). } 
\end{figure}

The electronic contribution to the free energy of the
junctions can be written in terms of the ABS energies
\beq
F_{el} = -k_B{T}\sum_{\bf{k}}\sum_{\sigma} \frac{|k_{z}|}{k_{F}}
\log\left[2\cosh\left(\frac{E_{{\bf{k}},\sigma}}{2k_B{T}}\right)\right] \label{eq:Fel}   
\eeq
As in~\eq{eq:GL}, the magnetic free energy of the barrier is included to
lowest-order $F_{mag}=M^2/2\chi$, where $\chi$ is given in units of $(g\mu_{B}/\hbar{v_F})^2/\Delta_{0}$. Numerically minimizing the total free energy
$F_{el}+F_{mag}$ with respect to both $M$ and
$\phi$, we find the global free energy minimum. Typical minimizing values
for the \pzpz and \pypy junctions are plotted as a function of temperature
in~\fig{fig3}(a) and~\fig{fig3}(b) respectively. 
We find  that the barrier undergoes a
magnetic instability  
at sufficiently large $\chi$, and below a critical temperature $T_{M}
< T_c $ such that the 
magnetic state appears only in the presence of superconductivity. For 
$Z\neq0$ the junction is in a fractional state below $T_{M}$ with two
degenerate free energy minima
$(M,\phi_{\text{min}})$ and $(-M,-\phi_{\text{min}})$ (broken TRS). 

In~\fig{fig3}(c) and~\fig{fig3}(d) we show the phase diagram as a function of
$Z$ and $T$ at 
fixed $\chi$ for the \pzpz and \pypy junctions respectively. The qualitatively
different form of these phase diagrams follows from 
the response of the ABS spectrum to the appearance of the
magnetization and the resulting change in the
${\cal{T}}_{\sigma}({\bf{k}})$. As can be 
seen from~\eq{eq:ABSpzpz} and~\eq{eq:ABSpypy}, the 
ABS spectrum in the two junctions has very different
dependence upon ${\cal{T}}_{\sigma}({\bf{k}})$: due to the zero-energy states
in the \pzpz junction, each bound state $E_{{\bf k},\sigma}$ monotonically
shifts towards the middle of the gap with decreasing
${\cal{T}}_{\sigma}({\bf{k}})$; for the \pypy junction, in contrast, the
states move towards the gap edges. From~\eq{eq:Fel}, the free energy
contributed by $E_{{\bf{k}},\sigma}$ 
in a \pzpz junction will therefore increase
(decrease) as the transparency ${\cal{T}}_\sigma({\bf{k}})$ decreases
(increases), while the opposite is true for the \pypy junction. 

It immediately follows that the \pzpz junction is non-magnetic at
$Z=0$, as $M\neq0$ would reduce the transparency for 
both spin orientations and hence raise the total free energy. The $M=0$
state remains stable below some critical value of $Z$; increasing $Z$
beyond this, a magnetic moment appears as the free energy gain from decreasing
the transparency in the $\sigma=-1$ sector outweighs the increase in the
$\sigma=+1$ sector. Although the decrease in electronic free energy favors the
indefinite 
growth of ${M}$ with increasing $Z$, the maximum magnitude of ${M}$ is
limited by the cost in magnetic free energy. For 
sufficiently large $\chi$ and low temperatures, the transition back into the
non-magnetic state is first order, with regions in the phase diagram where the
non-magnetic and magnetic states are metastable, as shown in~\fig{fig3}(c). At
higher temperatures or smaller $\chi$, $|M|$ continuously
vanishes after going through a maximum.

In contrast, the \pypy junction displays a spontaneous magnetization at
$Z=0$ for all $T<T_{c}$, stabilized due to the reduction in $F_{el}$ from
the decreased transparency in each spin sector. This
effect is absent from the Ginzburg-Landau expansion of $F_{el}$ in~\eq{eq:GL},
as we have only kept terms to first order in ${\bf{M}}$; from~\eq{eq:ABSpypy},
however, we see that the magnetization only enters $F_{el}$ as
$|{\bf{M}}|^2$ when $Z=0$. Despite the spontaneous magnetization, the
junction is not in a fractional state and $\phi_{\text{min}}=0$.
Turning on a finite tunneling barrier strength ($Z>0$) at fixed $T$, the
magnetic moment of the barrier 
decreases to compensate for the free-energy increase from the
enhanced transparency in the $\sigma=+1$ sector; $\phi_{\text{min}}$
simultaneously takes on a fractional value. As $Z$ is further
increased, the barrier moment is monotonically suppressed, while the stable
phase difference passes through a stationary point before returning to its 
$Z=0$ value.

\begin{figure}
  \includegraphics[width=0.8\columnwidth]{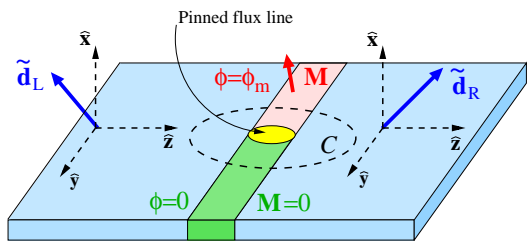}
  \caption{\label{fluxline} (color online) Proposed experiment for the
    observation of fractional flux quanta at the interface between magnetic
    and non-magnetic regions of a tunneling barrier. The contour ${\cal{C}}$
    is used in evaluating~\eq{eq:flux}.}   
\end{figure}

The characteristic signature of the magnetic instability is the appearance of
fractional flux quanta at the 
junction interface. In~\fig{fluxline} we show a proposal for their observation
in a Josephson junction  
with a tunneling barrier consisting
of two materials, one of which undergoes the magnetic
instability proposed here, while the other remains non-magnetic at all
temperatures. The stable phase difference across the magnetic region is   
$\widetilde{\phi}$, while it is $0$ across the non-magnetic region. If a
magnetic flux line is trapped at the boundary between the barrier segments,
a line integral along the contour ${\cal{C}}$ 
in~\fig{fluxline} shows that the enclosed flux $\Phi$ is  
\beq
\frac{\Phi}{\Phi_{0}} = n + \oint_{{\cal{C}}}d{\bf{s}}\cdot\nabla\phi
= n + \frac{\widetilde{\phi}}{2\pi}, \qquad n\in{\mathbb{Z}} \label{eq:flux}  
\eeq 
As the temperature is lowered below $T_{M}$, $\widetilde{\phi}$ takes a
fractional 
value, and a flux line appears with a continuously increasing flux
$\Phi \neq \Phi_0 $. Experimentally, such a flux line could be
directly observed by local magnetic probes like scanning SQUID 
microscopy, or inferred from the asymmetric Fraunhofer pattern of critical
current vs applied field. This proposal resembles 
the devices used to observe half-integer flux quanta by Weides \emph{et
  al.},~\cite{WeidesPRL2006} where spin-singlet 
superconductors were used instead of spin-triplet, and differing widths of a
permanent magnetic barrier guaranteed $\widetilde{\phi}=\pi$ always. 
Although other proposals exist for the creation of fractional flux
quanta,~\cite{Sigrist1995,GoldobinPRB2004} their detection in
our proposed junction would be unambiguous confirmation  
of the magnetic instability.

\begin{figure}
  \includegraphics[width=\columnwidth]{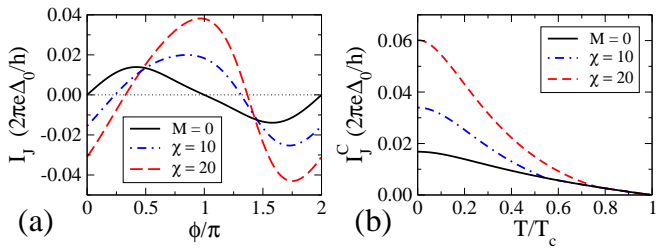}
  \caption{\label{current} (color online) (a) Current vs phase relationships
    in the \pypy junction at $T=0.2T_{c}$ both with and without ($M=0$) the
    magnetic instability. (b) Critical Josephson current as a function of
    reduced temperature in the \pypy junction. In both panels we take
    $\eta=0.2\pi$ and $Z=0.7$.}
\end{figure}

The magnetic instability of the tunneling barrier radically alters the
supercurrent transmission through the junction. From~\eq{eq:GL}, we find that
the Josephson current vs phase relationship 
$I_J=(e/\hbar)\partial{F}/\partial{\phi}\propto\sin(\phi-\phi_{\text{min}})$
is
shifted from its usual form for a non-magnetic barrier. This is clearly seen
in~\fig{current}(a) for the \pypy junction (the \pzpz junction results are
qualitatively identical). Note that the different magnitudes of $\max\{I_{J}\}$
and $\min\{I_{J}\}$ are due to higher-order harmonics in $\phi$ which are not
included in the free energy
expansion~\eq{eq:GL}. We also find a strong enhancement of the critical current
$I^{C}_{J}=\max\{|I_{J}|\}$ below the magnetic instability [\fig{current}(b)].
This occurs as the increased current through the spin sector with the enhanced
transparency over-compensates for the decreased current through the spin
sector with the lowered transparency. The increase of $I^{C}_{J}$ below
$T_{M}$ is
reminiscent of the ``low-temperature-anomaly'' of $d$-wave Josephson
junctions.~\cite{Kashiwaya2000}

For simplicity, we have neglected the suppression of the TSC state 
near the interface due to the proximity effect. Including this
would increase $F_{el}$ and hence shrink the parameter space where the moment
is stable. Although quantitative 
changes in the phase diagrams~\fig{fig3}(c) and (d) are expected, the free
energy expansion~\eq{eq:GL} and the basic experimental signatures of the
fractional state remain valid. 
Furthermore, we do not expect a qualitatively different ABS spectrum, and so
the essential role of the 
orbital pairing state of the two TSCs should remain in a fully
self-consistent analysis.

In conclusion, we have shown that a TRS-breaking configuration of two TSCs in 
a Josephson junction can cause the tunneling
barrier to develop a spontaneous magnetization. This realizes an exotic
Josephson state with stable phase difference 
$0<\phi_{\text{min}}<\pi$. The orbital part of the TSC pairing state was
demonstrated to control the magnetic instability. The existence of fractional
flux quanta at the barrier, and a large increase in the critical current
beneath the magnetic transition temperature, are the experimental signatures
of this state.

PMRB thanks the Center for Theoretical Studies of ETH Zurich and the MPI-FKF
in Stuttgart for their kind hospitality. We are also grateful to Y. Asano,
B. Hamprecht, Y. Maeno and J. Sirker for helpful discussions.

\end{document}